\documentclass[nofootinbib,twocolumn,prc,aps]{revtex4-1}
\usepackage{hyperref}
\usepackage{enumerate}
\usepackage{amsmath}
\usepackage{amsfonts}
\usepackage{amsthm}
\usepackage[usenames,dvipsnames]{color}
\usepackage{graphicx}
\usepackage{epstopdf}

\newcommand{\der}[2]{\frac{\partial #1}{\partial #2}}
 
\newcommand{\derc}[3]{\displaystyle \left( \frac{\partial #1}{\partial #2} \right)
\raisebox{-1em}{\ensuremath{#3}}}

\def\dublet#1#2{\left(\raisebox{6.5pt}{\it #1}\llap{\raisebox{-4.5pt}{\it #2}}
\,\right)}

\def\dublet#1#2{\left(\raisebox{6.5pt}{\it #1}\llap{\raisebox{-4.5pt}{\it #2}}
\,\right)}

\graphicspath{{figs/}}

\begin{document}
 
\title{Influence of the interactions of scalar mesons on the behavior of the symmetry energy}

\author{Noemi Zabari}
\affiliation{Henryk Niewodnicza\'nski Institute of Nuclear Physics,\\
Polish Academy of Sciences,\\
ul.Radzikowskiego 152, 31-342 Krak\'ow, Poland}
\author{Sebastian Kubis}
\email{skubis@pk.edu.pl}
\author{W\l odzimierz~W\'ojcik}
\address{Institute of Physics, Cracow University of Technology,  Podchor\c{a}\.zych 1, 
30-084 Krak\'ow, Poland}

\begin{abstract}
Symmetry energy behavior of scalar mesons interactions is analyzed within the framework of the standard 
relativistic mean field model.
Whereas the presence of the $\delta$ meson itself makes the symmetry energy stiffer, the crossing term 
$\delta\!-\!\sigma$  allows its slope to decrease to the suggested experimental value.
Moreover, such controlling of the symmetry energy does not significantly affect the stiffness of the equation of state and acceptable neutron star masses result. Interestingly, for the most plausible value of the symmetry energy slope, the phase transition occurs in the neutron star core.
\end{abstract}

\maketitle

%% abstrakt niepoprawiony !!

\section{Introduction}

One of the most relevant parameters in the study of nuclear matter is the nuclear symmetry energy $E_{sym}$. 
It describes the increase of energy with asymmetry of matter. 
More precisely, the symmetry energy is defined as the second derivative of the total energy with respect to the matter asymmetry at a given density $n$. Hence, it is a function of $n$.
 This density dependence is currently the  subject of  intense  experimental and theoretical research.
It is commonly accepted that at the saturation density $n_0=0.16 \rm ~fm^{-3}$, the symmetry energy value 
$J=E_{sym}(n_0)$ is approximately  30~MeV. 
 In the last decades, great progress has been made in the determination of the symmetry energy behavior around
 $n_0$. 
The most important parameter in this field of study is the symmetry energy slope $L$ defined as the symmetry energy derivative 
\begin{equation}
  L=3 n_0 \left.\der{E_{sym}}{n}\right|_{n_0} ~.
\end{equation}  
For a long time, the value of this slope has been a matter of discussion due to inconsistent  experimental results.  
Early studies based on isospin scaling  suggested  $L$ to be well above 100~MeV 
\cite{Chen:2004si}. The neutron skin thickness measurement \cite{Chen:2010qx} resulted in much lower $L$, 
of around 65~MeV. Although the slope is a simply defined quantity, it is difficult to extract its value experimentally.
  A review \cite{Li:2013ola} of a few years ago,  presented  a variety of approaches for the determination of $L$.
  Given values were between  40 and 70 MeV,  including uncertainties  the range becomes even larger: between 20 and 120 MeV.
 In other  work \cite{Lattimer:2012xj}, the authors analyzed the correlation between $J$ and $L$
 with constraints arising from different experiments and astrophysical observations.
 They concluded that the slope lies in the range 40.5 - 61.9 MeV at a 90\% 
 confidence level. 
 One might expect that the ongoing experimental progress will narrow the range of possible values of $L$.
 The most recent review \cite{Oertel:2016bki} collected results from all relevant experiments and astrophysical observations and concluded that an acceptable value for the slope is $L=58.7\pm 28.1~\rm MeV$.

 Furthermore, theoretical studies on the symmetry energy slope also gave a relatively wide range of possible values of $L$.
A well-known tool used for the description of dense matter is the Relativistic Mean Field (RMF) theory
\cite{Walecka:1974qa}.
It is constrained by experimental results at saturation point density and it allows extrapolation of the nuclear matter parameters to the higher or lower densities essential in the analysis of neutron star structure.
In its basic form, the RMF model predicts a rather high $L$, of up to 140 MeV even. Only its modified form: 
DDRMF (density-dependent RMF - where the coupling constants are functions of density) allows one to obtain 
acceptable values of the symmetry energy slope \cite{Typel:2005ba}.
For comparison of various RMF models in the context of the symmetry energy see Ref.~\cite{Chen:2007ih}.

However, the description of nuclear matter based on Brueckner-Hartree-Fock theory or the Skyrme model gives a very broad range for $L$, dependent on the parameters used in the model. A broad discussion of various models is presented in this review \cite{Li:2008gp}.

The standard RMF model can be understood as a minimalistic construction based on the introduction of Yukawa coupling of nucleons to mesons where the coupling constants are independent of the density. 
The model only includes meson fields which couple to all possible nucleonic currents:$\bar{\psi}\psi,~~ 
\bar{\psi}\gamma^\mu\psi, ~~ \bar{\psi}\vec{\tau}\psi,~~ \bar{\psi}\gamma^\mu\vec{\tau}\psi, ~~ \bar{\psi}\psi$. Hence we get $\sigma, \omega, \delta$ and $\rho$ mesons, respectively. 
The $\deltaδ$ meson can be said to be negligible as its influence on nuclei properties is not large, so the symmetry energy can be expressed by 
  \begin{equation}
    E_{sym} = \frac{k^2_0}{6 E_0} + \frac{g_\rho^2}{8 m_\rho^2} n~ ,
    \label{Esym-rho}
  \end{equation}
where $k_0$ and  $E_0$ are the Fermi momentum and energy for symmetric matter respectively and
 $g_\rho, m_\rho$ 
are the coupling constant and mass of the $\rho$ meson respectively.
The first term in Eq.(\ref{Esym-rho}) is from the difference of the Fermi levels of the protons and neutrons, the second  comes from the nucleon-$\rho$ interaction.   
The second term dominates the first (which scales approximately as $n^{1/3}$) and causes a high value of 
$L$. 
Different  work \cite{Horowitz:2001ya} showed that the rapid growth of $E_{sym}$ due to the  $\rho$
 meson could be diminished by inclusion of meson crossing terms $\rho\!-\!\sigma$   and $\rho\!-\!\omega$.
These terms increase the effective mass $m_\rho$  and suppress the second term in Eq.(\ref{Esym-rho}).
 Effects of the $\rho\!-\!\omega$ crossing term in the context of neutron stars were explored 
 \cite{ToddRutel:2005fa,Fattoyev:2010rx}, where it appeared that the obtained equation of state was too soft (the maximal neutron star mass obtained was around $1.7 M_\odot$).
The model was later improved \cite{Fattoyev:2010mx,Cavagnoli:2011ft} and a higher mass obtained 
$1.97 M_\odot$, but this was still too low compared to the most recent observations  \cite{maxmass1,maxmass2}.
Besides the $\rho$ meson in the isovector sector, the scalar meson $\delta$  should also be included, as was proposed in Ref.~\cite{Kubis:1997ew}.
  It was discussed there that although the $\delta$ meson contribution to the total energy is negative, its inclusion makes the symmetry energy increase due to the vector meson $\rho$ contribution which dominates all other terms and in the end makes $L$ large.
In this work, we propose an extension to the standard RMF approach, namely studying the interaction of the scalar-isoscalar meson $\sigma$  with the scalar-isovector  $\delta$  to analyze the symmetry energy slope and its behavior at higher densities. 
Scalar meson interactions appear naturally in chiral perturbation theory (ChPT), where the chiral expansion introduces various meson-meson vertices of different powers \cite{Furnstahl:1996wv}. 
 ChPT extended to $SU(3)_L\times SU(3)_R$ symmetry has been successfully applied to describe nuclei  \cite{Papazoglou:1998vr}. 
Subsequently, this approach was applied to asymmetric matter relevant for neutron stars  
\cite{Dexheimer:2008ax,Dexheimer:2015qha}.
 It is worth noting that the lagrangian used in some works \cite{Dexheimer:2008ax,Dexheimer:2015qha}
 had to be extended to the isovector-scalar field ($\delta$-meson) to obtain correct properties of the 
symmetry energy. 
In this work, we consider the $\sigma\!-\!\delta$ interaction in the framework of RMF.
Both $\sigma$ and $\delta$ are responsible for attraction between nucleons, so their coupling could control 
the density dependence behavior of the symmetry energy. Furthermore, the $\sigma\!-\!\delta$
 coupling does not cause a large softening of the EOS as has happened with vector meson crossing terms 
 \cite{ToddRutel:2005fa,Fattoyev:2010rx}
 The simplest forms of the $\sigma$ and $\delta$ interaction, acceptable by isospin symmetry principles, are $\sigma \vec{\delta}^2$ or  $\sigma^2 \vec{\delta}^2$. Both of them have been taken into account and compared in this work.

\section{Formalism} 

We introduce a Lagrangian $\mathcal{L}$ which describes the particle interactions.
 The Lagrangian $\mathcal{L}$  contains
fields describing nucleons  $\psi=\dublet{$\psi_p$}{$\psi_n$}$  and four mesons
which are: the scalar-isoscalar $\sigma$ meson, vector-isoscalar $\omega$ meson, vector-isovector $\rho$ meson, scalar-isovector $\delta$ meson and additionally, the interaction between $\sigma$ and $\delta$
mesons.
The Lagrangian density $\mathcal{L}$ for nucleon and meson fields is given in
Refs.~\cite{Walecka:1974qa,Kubis:1997ew}:
\begin{widetext}
\begin{align} \label{lagrangian}
\begin{split}
\mathcal{L}& = 
\frac{1}{2}\left(\partial_\mu\sigma\partial^\mu\sigma-m_\sigma^2\sigma^2\right) -
\frac{1}{4}\left(\partial_\mu\omega_\nu-\partial_\nu\omega_\mu\right)\left(\partial^\mu\omega^\nu-\partial^\nu\omega^\mu\right)+\frac{1}{2}m_\omega^2\omega_\mu\omega^\mu  -
\frac{1}{4}\left(\partial_\mu{\vec{\rho}}_\nu-\partial_\nu{\vec{\rho}}_\mu\right)\left(\partial^\mu{\vec{\rho}}^\nu-\partial^\nu{\vec{\rho}}^\mu\right)+ \frac{1}{2}m_\rho^2{\vec{\rho}}_\mu{\vec{\rho}}^\mu
\\
 &+\frac{1}{2}\partial_\mu\vec{\delta}\partial^\mu\vec{\delta}-\frac{1}{2}m_\delta^2{\vec{\delta}}^2+\bar{\psi}\left(i\partial_\mu\gamma^\mu-m\right)\psi+g_\sigma\sigma\bar{\psi}\psi -g_\omega\omega_\mu\bar{\psi}\gamma^\mu\psi - \frac{1}{2}g_{\rho}\vec{\rho}_{\mu}\bar{\psi}\gamma^\mu\vec{\tau}\psi+g_\delta\vec{\delta}\bar{\psi}\vec{\tau}\psi-U(\sigma)  - 
{\cal L}_{\sigma \delta}, 
\end{split}
\end{align}
\end{widetext}
where $m$ is the nucleon mass, and $m_i$ are meson masses and, $i\!=\!\sigma, \omega, \rho$ and $\delta$ 
respectively. 
Here, $g_\sigma, g_\omega, g_\rho$ and $g_\delta$  are coupling constants for nucleons to the corresponding 
mesons and the potential $U(\sigma)$  is the self-interaction of the $\sigma$ meson that can be written as  
$U\left(\sigma\right)=\frac{1}{3}b \ m {(g_\sigma\sigma)}^3+\frac{1}{4}c\ {(g_\sigma\sigma)}^4$,
where $b$ and $c$ are dimensionless.
Such a potential is 
necessary to get proper compressibility value at the saturation point \cite{Boguta:1977xi}.   
The last term represents the $\sigma\textrm{-}\delta$ interaction and can  take two different forms 
\begin{equation}
  {\cal L}_{\sigma\delta} = \left\{ 
  \begin{array}{ll}
    \tilde{g}_{\sigma \delta^2}\sigma \vec{\delta}^2&   \rm linear ,\\
    \tilde{g}_{\sigma^2 \delta^2}\sigma^2 \vec{\delta}^2 &  \rm quadratic . \\ 
  \end{array}
  \right.
\end{equation}  
The field equations for the meson mean fields are as follows:
\begin{equation} \label{rown1}
m_\sigma^2\bar{\sigma}=g_\sigma\left(n_p^s+n_n^s\right)-\frac{\partial U}{\partial\sigma} -\tilde{g}_{\alpha}
\alpha\,\bar{\sigma}^{\alpha-1}\bar{\delta}^2 ,
\end{equation}
\begin{equation} \label{rown2}
m_\omega^2\ {\bar{\omega}}_0=g_\omega n ,
\end{equation}
\begin{equation} \label{rown3}
m_\rho^2{\bar{\rho}}_{03}=\frac{1}{2}g_\rho\left(2x-1\right)n ,
\end{equation}
\begin{equation} \label{rown4}
m_\delta^2{\bar{\delta}}_3=g_\delta\left(n_p^s-n_n^s\right) - 2\tilde{g}_{\alpha}\bar{\sigma}^{\alpha}\bar{\delta} ,
\end{equation}
where the $\sigma\!-\!\delta$ interaction terms include coupling constant depending on the 
type of interaction
\begin{equation} \label{g-int-razem}
  \tilde{g}_{\alpha}  = \left\{ 
  \begin{array}{ll}
   \tilde{g}_{\sigma \delta^2} ~~{\rm for~ linear,} ~ \alpha=1  ,\\
   \tilde{g}_{ \sigma^2 \delta^2} ~~{\rm for~ quadratic,} ~ \alpha=2 .  \\
  \end{array}
  \right.
\end{equation}  
In the framework of the RMF theory the mean values of meson the fields $\bar{\sigma}, \bar{\omega},
\bar{\rho},\bar{\delta}$ are determined by the vector and scalar nucleon densities $n_i$ and 
$n_i^s,~~i=n,p$.
In Eqs. (\ref{rown2}) and (\ref{rown3}), $n$ denotes the baryon density 
\begin{equation}\label{density}
  n=n_p+n_n=\frac{2}{{(2\pi)}^3}\int_{0}^{k_p}d^3k+\frac{2}{{(2\pi)}^3}\int_{0}^{k_n}d^3k
\end{equation}
 and $x$ is the number of protons per baryon  $x=\frac{n_p}{n}$.
Similarly, in Eqs. (\ref{rown1}) and (\ref{rown4}), $n_i^s$ is the proton and neutron scalar density respectively which has the form 
\begin{equation}\label{scalar-density}
 n_i^s=\frac{2}{{(2\pi)}^3}\int_{0}^{k_i}\frac{m_i^\ast}{\sqrt{k^2+m_i^{\ast2}}}d^3k ~~,~~~ i= p,n  
\end{equation}
where 
$m_p^\ast, m_n^\ast$  are the effective masses of the proton and neutron:
\begin{align}
 \label{meff-proton}
m_p^\ast=m-g_\sigma\bar{\sigma}-g_\delta{\bar{\delta}}_3 ,\\
\label{meff-neutron}
m_n^\ast=m-g_\sigma\bar{\sigma}+g_\delta{\bar{\delta}}_3 .
\end{align}
In the following, we use Fermi momenta for protons and neutrons 
 $k_i\!=\!\left(3\pi^2n_i\right)^\frac{1}{3},  i=n,p$ . 
The RMF approach allows one to present the  energy density
 $\epsilon(\bar{\sigma}, \bar{\omega}, \bar{\rho}, \bar{\delta}, 
k_p, k_n)$ as a function of mean field values and Fermi momenta. Due to 
 Eqs.~(\ref{meff-proton} and \ref{meff-neutron}) the energy density may be expressed in terms of nucleon 
 effective masses and densities
\begin{align} \label{energy-density}
\begin{split}
\epsilon  =& \frac{2}{(2 \pi)^3} \left( \int_{0}^{k_p} d^3 k \sqrt{k^2 + m_p^{\ast 2}} + \int_{0}^{k_n} d^3 k \sqrt{k^2 + m_n^{\ast 2}} \right) \\&
+ \frac{1}{2} \frac{1}{C_{\sigma}^2} (m - \bar{m}^*)^2 + 
 \frac{1}{2} C_{\omega}^2 n^2 + \frac{1}{8} C_{\rho}^2 (2x-1)^2 n^2 \\ 
 & +\frac{1}{8} \frac{1}{C_{\delta}^2} (\Delta m^*)^2  
 { \; + \; g_{\alpha} (\Delta m^*)^2 (m - \bar{m}^*)^\alpha } 
+ U(m - \bar{m}^*),
\end{split}
\end{align}
where $\bar{m}^* = (m_p^* + m_n^*)/2$ and $\Delta m = m_n^* - m_p^*$.
In the expression for the energy, we introduce $g_\alpha$ instead of $\tilde{g}_\alpha$, i.e. 
$ g_\alpha = \displaystyle\frac{\tilde{g}_\alpha}{4 g_\sigma^\alpha g_\delta^2}$. 
We use $g_\alpha$ for convenience in further proceeding calculations.
One must remember that the parameter $g_\alpha$ takes a different form in the quadratic or linear case and has different units.
The translational invariance of the meson mean fields and the use of effective masses in the energy density function allows one to replace the coupling constants $g_\sigma, g_\omega, g_\rho, g_\delta$ by
$C^2_\sigma=\frac{g_{\sigma}^2}{m_{\sigma}^2},\ C^2_\omega=\frac{g_{\omega}^2}{m_{\omega}^2},\ C^2_\rho=\frac{g_{\rho}^2}{m_{\rho}^2},\ C^2_\delta=\frac{g_{\delta}^2}{m_{\delta}^2}$
 and hereafter we treat them as the relevant model parameters.

\section{RESULTS}

The seven coupling constants: $C_i^2, b,c$ and $g_\alpha$  - have to be adjusted to fit the saturation 
properties of nuclear matter. 
$C^2_\sigma, C^2_\omega, b, c$ belong to the isoscalar sector and the remaining three $C^2_\rho, 
C^2_\delta, g_\alpha$ belong to the isovector sector.
 Using experimental saturation properties such as a saturation density $n_0=0.16\ \rm {fm}^{-3}$, the binding energy for symmetric nuclear matter	
 $B = \epsilon\left(n_0, \frac{1}{2}\right)/n_0 -m = -16\ \rm MeV$
 and an incompressibility $K=230\ \rm MeV$, we are able to extract only three isoscalar coupling constants: $\ C_\omega^2=6.48\ \rm fm^2$, $ b=0.0054$ and $c=-0.0057$ which means that $C_\sigma$ remains undetermined.
To some extent, it may be restricted from the stiffness of the equation of state, namely the pressure versus
density relation. It was shown by Prakash {\em et al.} \cite{Prakash:1988md}, that due to the 
correlation between scalar and  vector meson couplings,
the greater $C_\sigma^2$ is, the greater the pressure at densities above  $n_0$. 
So, the higher $C_\sigma^2$ is, the stiffer the equation of state (EOS) becomes.
In this work, we employ the coupling constant $\ C_\sigma^2 =11\ \rm fm^2$ 
which gives the EOS sufficient stiffness  to obtain neutron star mass above  $2 M_\odot$.
Summing up, the following saturation properties: $n_0,B,K$ and the stiffness of the EOS determine
coupling constants  from isoscalar sector. 

The set of parameters describing the interactions of isovector mesons, 
appearing in Eq. (\ref{energy-density}), $\ C_\rho^2 $, $\ C_\delta^2$ and $g_{\alpha}$, are strictly connected to the symmetry energy $E_{sym}(n)$. 
The symmetry energy is defined as the second derivative of the energy density $\epsilon\left(n,x\right)$:
\begin{equation} \label{Esym-def}
E_{sym}(n)=\left.\frac{1}{8n}\frac{\partial^2\epsilon\left(n,x\right)}{{\partial x}^2}\right|_{x=\frac{1}{2}}.
\end{equation}
From Eq.~(\ref{Esym-def}), we obtain the symmetry energy in the following form:
\begin{align} \label{Esym-eq}
\begin{split}
E_{sym} & (n)=\frac{1}{8} C_{\rho}^2 n + \frac{k_0^2}{6 \sqrt{k_0^2 + m_0^{*2}}} \\ 
 &- C_{\delta}^2 \frac{m_0^{*2} n}{2 (k_0^2 +m_0^{*2})\left( 1 + C_{\delta}^2 A + 8\,  g_{\alpha} C_{\delta}^2\, (m - m_0^*)^\alpha \right)},
\end{split}
\end{align}
\begin{figure}[t]
\begin{center}
\includegraphics[width=1\columnwidth]{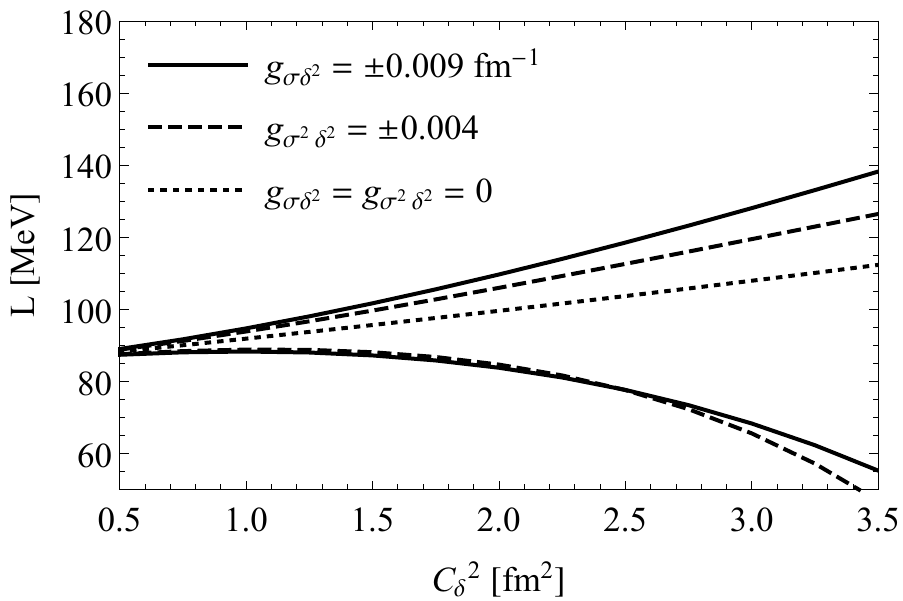}
\caption{ Slope of symmetry energy $L \ \rm \left[MeV\right]$ as a function of $C_\delta^2 \ \rm [fm^2]$ for different 
$ g_\alpha$. For negative $ g_\alpha$ the slope is decreasing.}
\label{fig:plotSlope}
\end{center}
\end{figure}
where $A(k_0, m_0) = \frac{4}{(2 \pi)^3} \int_{0}^{k_0}  \frac{k^2}{(k^2 +m_0^{*2})^{3/2}}  d^3\!k$. For symmetric matter at saturation density, the effective  masses and Fermi momenta are equal, hence $m_p^{\ast} = m_n^{\ast} = m_0^{\ast}$ and $k_p= k_n = k_0$ where $k_0(n)={(\frac{3}{2}\pi^2n)}^{1/3}$ .
The coupling constants for $\rho$ and $\delta$ mesons can be correlated in such a way as to fix the symmetry energy $E_{sym}(n_0)$. Here we adopt $E_{sym}(n_0) = 30 ~\rm MeV$. A detailed
discussion can be found in Ref.~\cite{Kubis:1997ew}.
We have extended this formalism to include the $\sigma\textrm{-}\delta$ crossing term with its coupling determined by  $g_{\alpha}$ to control the value of the symmetry energy slope $L$.
While the value of $E_{sym} (n_0)$ is determined to be $30\ \rm MeV$, we allow the slope $L$ to scatter
over a broad range. Then, the free parameters  $C_{\delta}^2$ and $g_{\alpha}$, were constrained in such a way that the slope $L$ varies in the range from $48$ to $140 ~\rm MeV$. Values adopted for further calculation are shown in Table \ref{table2}.
\begin{table}[b!]
\caption{Parameters in isovector sector.}
~\\
\label{table2}
\begin{tabular}{|l||c|r|l|c|r|l|} 
\cline{2-5} 
 \multicolumn{1}{c}{} & \multicolumn{2}{|c}{Linear} \vline  &  \multicolumn{2}{c}{Quadratic} \vline\\
\cline{2-5} 
\multicolumn{1}{c}{} & \multicolumn{2}{|c}{$g_{ \sigma \delta^2} = 0.009~ \rm fm^{-1}$ } \vline  &  \multicolumn{2}{c}{$g_{ \sigma^2 \delta^2} = 0.004$} \vline\\ \hline
  $C_{\delta}^2 ~\rm [fm^2]$  & $C_{\rho}^2~\rm [fm^2] $ & $L~ \rm [MeV]$   &   $C_{\rho}^2~ \rm [fm^2] $ & $L~ \rm [MeV]$ \vline \\
  $0.5$ & $5.2$ & $88.9$ & $5.3$ & $88.1$ \\
  $1.5$ & $8.2$  & $98.4$ & $8.4$  & $100.9$ \\
  $2.5$ & $10.6$ & $112.3$ & $11.3$ & $117.9$ \\
  $3.5$ & $12.7$ & $126.6$ & $13.8$ & $139.3$ \\
\cline{2-5} 
 & \multicolumn{2}{c}{$g_{ \sigma\delta^2} = - 0.009~ \rm fm^{-1}$ } \vline  &  \multicolumn{2}{c}{$g_{ \sigma^2\delta^2} = - 0.004$} \vline\\ \hline
   $C_{\delta}^2 ~\rm [fm^2]$  & $C_{\rho}^2~ \rm [fm^2] $ & $L ~\rm [MeV]$   &   $C_{\rho}^2~ \rm [fm^2] $ & $L ~\rm [MeV]$ \vline \\
  $0.5$ & $5.4$ & $87.9$ & $5.4$ & $86.8$ \\
  $1.5$ &  $9.6$ & $87.9$ & $9.3$ & $86.5$ \\
  $2.5$ & $14.8$ & $76.9$ & $13.6$ & $77.3$ \\
  $3.5$ & $21.3$ & $47.2$ & $18.4$ & $54.4$ \\ \hline  
\end{tabular}
\end{table}
In Fig. \ref{fig:plotSlope} we show the correlation between the symmetry energy slope and
 $\sigma\!-\!\delta$ coupling.
Here one can notice the importance of such a crossing term. 
Without the  $g_{\alpha}=0$ term, the value of the symmetry energy slope cannot be fully controlled.
A change in $C^2_\delta$  can only increase the slope and an $L$ above 90~MeV results, which is hardly 
acceptable (regarding the  experimental data).
Introduction of  $\sigma\!-\!\delta$  coupling allows for better control of the slope. 
Particularly, due to the negative coupling constant $g_{\alpha}$, the slope can be significantly
diminished.
\begin{figure}[t!]
\begin{center}
\includegraphics[width=\columnwidth]{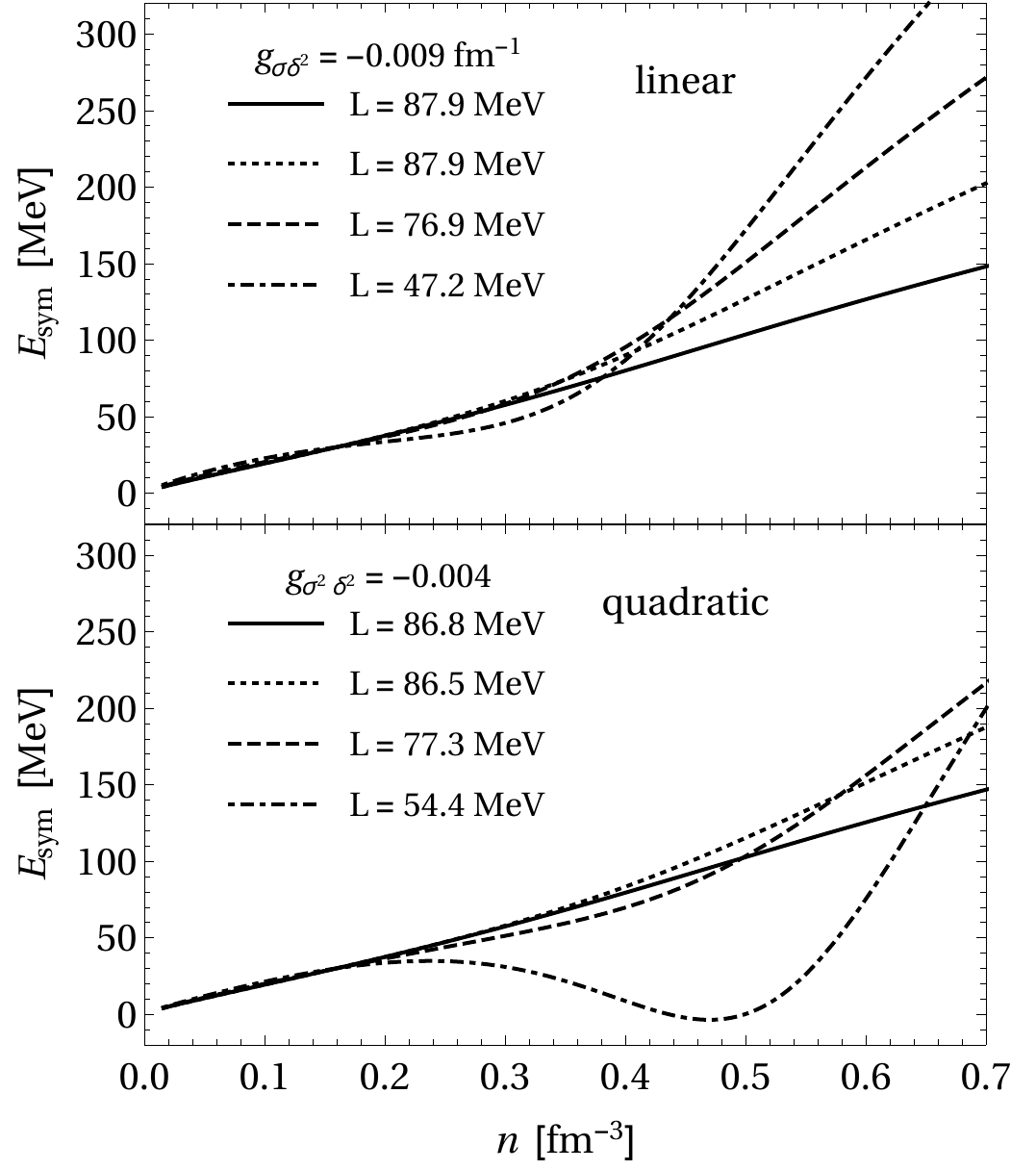}
\caption{ Symmetry energy $E_{sym}$ for negative value of $ g_\alpha$ with respect to various values of $C_\delta^2$ for both  types of coupling.}
\label{fig:plotEsym}
\end{center}
\end{figure}
The upper panel in Fig.~\ref{fig:plotEsym} shows the symmetry energy $E_{sym}$  in the case of linear 
coupling  where $g_{\sigma\delta^2}= -0.009 ~\rm fm^{-1}$  with different values of 
$C_\delta^2=0.5,1.5,2.5,3.5~ \rm fm^2$.  Corresponding values of $L$ are given in the plots. 
The sign of  $g_\alpha$ decides the role played by the  coupling constant $C_\delta^2$. 
For positive $ g_\alpha$, the slope $L$ attains unacceptable values and this also leads to the symmetry 
energy rapidly increasing with the density. For negative $g_\alpha$, the influence of 
$C_\delta^2$ is more complex.
Slightly above $n_0$, the increase of $E_s$  is small and at much higher densities it grows rapidly.
A similar tendency is observed for quadratic coupling but it is much more apparent, see the 
bottom panel in  Fig.~\ref{fig:plotEsym}.
It is interesting that in the quadratic case, for sufficiently high $C_\delta^2$, the symmetry energy becomes a decreasing function at some densities and even attains negative values.
%%%%%%%% new
Such super-soft symmetry energy has never been obtained in RMF models. However, this kind of $E_{sym}$ 
behavior is  quite common in approaches based on the MDI and Skyrme interactions \cite{Li:2008gp}, and more recently, it was also obtained in the framework of Nambu-Jona-Lasinio model \cite{Wei:2015aep}. 
%%%%%%%%%%%

\begin{figure}[b!]
\includegraphics[width=\columnwidth]{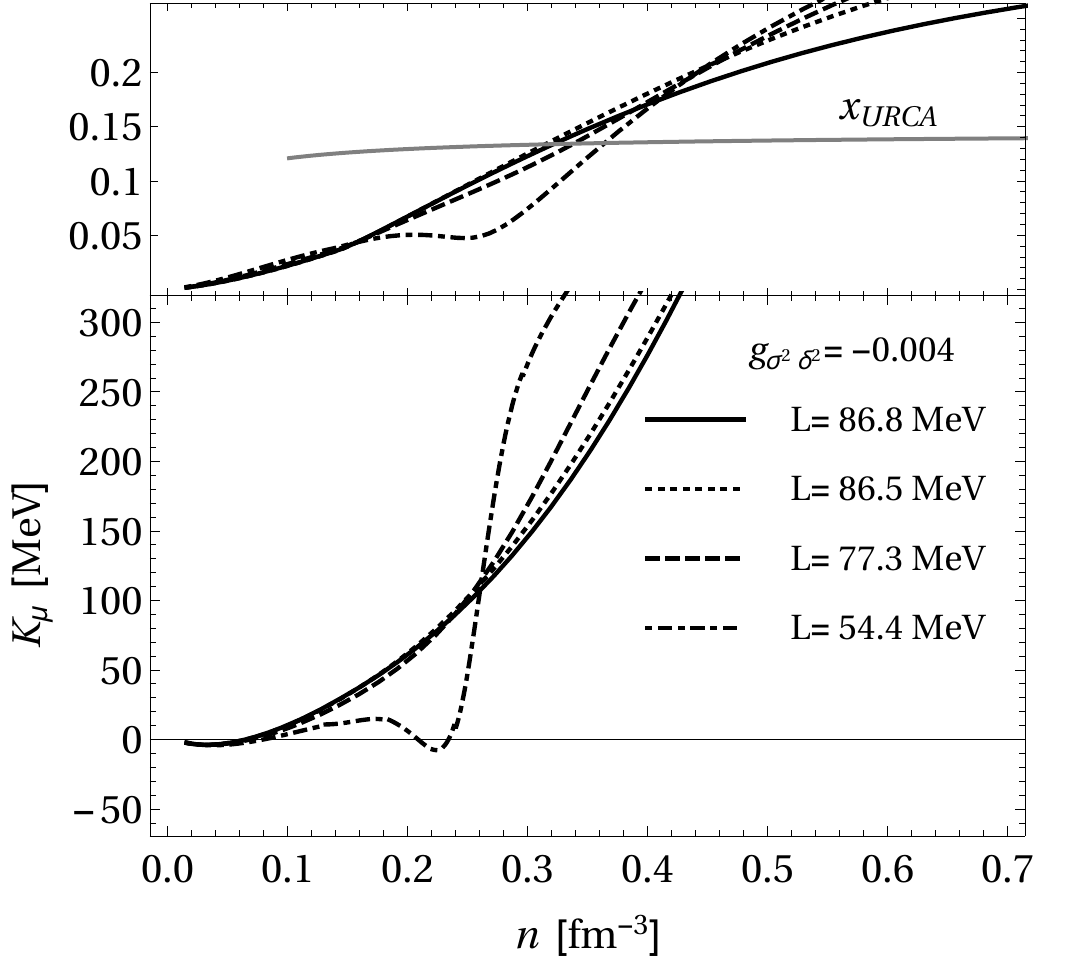}
  \caption{The proton fraction and  compressibility of  beta-equilibrated  matter for the quadratic model 
  with various $C_\delta^2$ }
\label{fig:compress}
\end{figure}

\begin{figure}[t!]
\begin{center}
\includegraphics[width=\columnwidth]{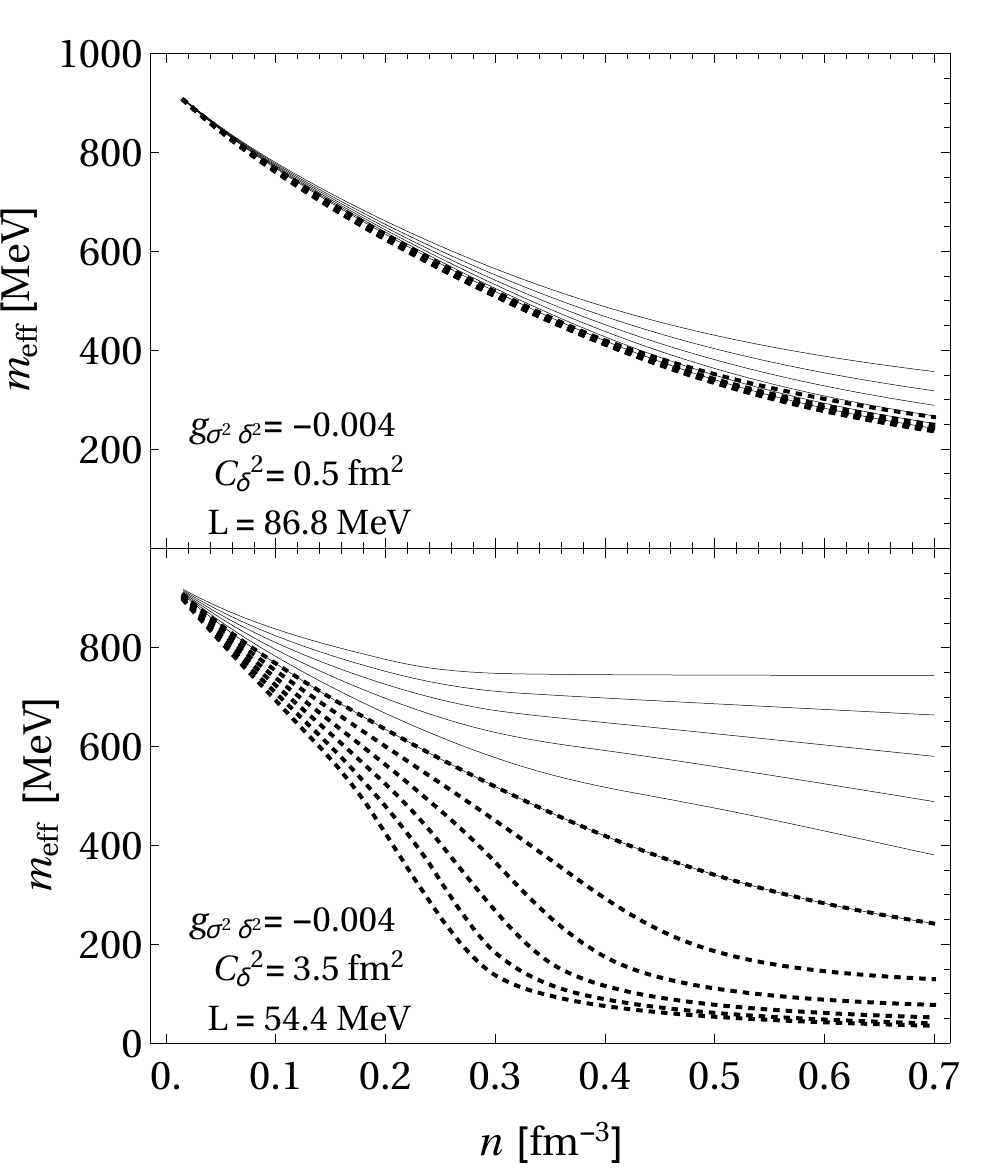}
\caption{ Effective masses of protons (solid lines) and neutrons (dashed lines) for two values of $C_\delta^2$ for quadratic type of coupling.}
\label{fig:plotMass}
\end{center}
\end{figure}

Such density regions of decreasing symmetry energy are interesting with respect to the stability of the neutron star matter.
It was shown in Ref.~\cite{Kubis:2006kb} that the symmetry energy is a crucial quantity determining whether the matter under beta equilibrium remains as a one-phase system or splits into two phases.
The signal of the phase separation comes from the negative value of the matter compressibility taken as the pressure derivative with respect to density under constant electron chemical potential $\mu$.
It can also come from the negative value of the charge susceptibility $-$ the derivative of charge with respect to $\mu$ under constant baryon density:
\begin{equation}
  K_\mu = \derc{P}{n}{\mu} >0 ~~,~~ \chi_n = \derc{q}{\mu}{n} >0 ~.
  \label{satbility}
\end{equation} 
 Indeed, with the highest $C_\delta$ coupling, where the slope $L$ assumes the most plausible value 
 ($\sim$50 MeV), the compressibility becomes negative, shown in Fig.~\ref{fig:compress}.
Such behavior of the compressibility at the density corresponding to the neutron star core is very interesting. It means that matter in the core cannot be homogeneous but forms a two-phase system. 
It is likely that some portion of the liquid core becomes solid as was suggested in 
Ref.~\cite{Kubis:2007zz}.
The presence of the $\delta$ mean field makes the effective nucleon mass splitting,$m_p$ and $m_n$ no longer the same. 
In Fig.~ \ref{fig:plotMass} we have depicted the effective masses of protons and neutrons as a function of baryon density  $n$ for various values of the proton fraction $x$ for the quadratic case.
Similar behavior is also observed for the linear term.
The solid lines correspond to proton masses $m_p$, while dashed lines correspond to neutron masses $m_n$.
The proton fraction ranges from $x= 0$ for pure neutron matter (edge lines) to $x=0.5$ for symmetric matter 
(middle line). The lines in between represent $x = 0.1,0.2,0.3 \text{ and } 0.4$. 
It is natural that the higher  $C_\delta^2$ is, the bigger the mass splitting becomes. 
For the most plausible value of $L$, the mass difference becomes very large - it is comparable to the rest 
mass of a nucleon.
Here we present the results for negative $\delta\!-\!\sigma$ coupling only, which is acceptable with respect to the symmetry energy slope.
Results for $g_{\sigma^2 \delta^2} > 0$, look similar. However, the effective mass splitting is smaller.

The proposed model must be confronted with the basic parameters of neutron stars. By solving the Tolman-Oppenheimer-Volkoff equations we acquire the family
of stars parametrized by the central density for the whole parameter space.
The essential quantity is the maximum mass of the stellar configuration.
For the linear model with $g_{\sigma\delta^2}=-0.009 ~\rm fm^{-1}$ the maximum mass, the radius of maximal configuration and its central density have the ranges: 
 $M_{max} = 2.17 - 2.21~ M_\odot$,~ $R_{max} = 11.09 - 11.53~\rm km$, and  $n_{cen} = 1.05 - 0.97~ \rm fm^{-3}$
  corresponding to the range of  $C_\delta^2 = 0.5 - 3.5~ \rm fm^2$.
In the quadratic model case, $g_{\sigma^2\delta^2}=-0.004$ and the considered range of couplings was slightly  smaller, which means that $C_\delta^2=0.5 - 2.5$.
Then the stellar  parameters take ranges $M_{max} = 2.17 - 2.22~ M_\odot$,~ $R_{max} = 11.09 - 11.37~\rm km$ 
and  $n_{cen} = 1.05 - 0.99~ \rm fm^{-3}$. 
The obtained maximum masses are in agreement with present observations \cite{maxmass1,maxmass2}.
For the most plausible values of $C_\delta^2=3.5 ~\rm fm^2$ (which gives a slope $L\sim$ 50 MeV) the relation density versus pressure is not unique because of the negative compressibility; see Fig.~\ref{fig:compress}.
 Such behavior requires a construction of a two-phase system. In this case, acquiring the proper equation of state presents a problem exceeding the scope of this paper, so it will be presented in a separate work.
 In Fig.~\ref{fig:compress} the proton fraction in the quadratic model for different $C_\delta$ couplings is shown (a similar dependence was obtained for the linear model). 
This quantity is relevant for neutron star thermal history.
When the proton fraction exceeds the threshold value for the so-called direct URCA cycle \cite{lattimer91} 
the star is cooled very efficiently, (provided no other effects, like superfluidity are present 
\cite{Yakovlev:2000jp}). 
In all considered  models the critical density for dURCA is not high: it ranges
 from 0.25 to  0.3 $\rm fm^{-3}$. 
This means that the critical neutron star mass for all cases takes values between 1.01 and 1.06  solar mass, well below the typical masses of observed neutron stars.    

\section{CONCLUSIONS}

In this paper we have shown that the inclusion of scalar meson coupling in the standard RMF model may be used 
to control the value of the symmetry energy slope $L$.
The relevance of the meson-meson interaction in the RMF model is detailed.
The slope $L$, consistent with the experimental constraints can be obtained both for linear or quadratic 
versions  of the $\sigma\textrm{-}\delta$  coupling only if the coupling constant takes negative values.
A difference between these two sorts of meson interactions manifests only at higher densities.
The quadratic coupling lowers the symmetry energy much more efficiently than the linear one.
In the former case, the symmetry energy can be close to zero.
Such behavior of the symmetry energy is unusual in the RMF approach and moreover has interesting consequences 
for the equation of state. 
At the densities where the symmetry energy vanishes, the compressibility of matter is negative which is the signal of the phase transition to the system with two different phases. This point requires further research. 
The quadratic coupling can also make the effective mass splitting very large. 
Already at a few $n_0$, the mass splitting is comparable to the nucleon rest mass.
For the models with positive compressibility, the equation of state was derived in order to check the basic 
neutron star properties. The obtained masses agree with observations.
With reference to the neutron star cooling, it appeared that the proton fraction required for fast cooling is attainable for stars with masses slightly above $1 M_\odot$.

\bibliography{delta-sigma}

\end{document}